\documentclass[final]{svjour2}
\usepackage{graphicx}
\usepackage{rotating}
\usepackage{amssymb}
\usepackage{mathptmx}
\usepackage[square, comma, sort&compress]{natbib}
\usepackage{caption}
\makeatletter
\journalname{Journal of Low Temperature Physics}

\bibpunct{}{}{,}{s}{}{,}

\begin{document}
\setcitestyle{numbers,square,sort&compress}

\newcommand{\hdblarrow}{H\makebox[0.9ex][l]{$\downdownarrows$}-}
\title{Design and Deployment of a Multichroic Polarimeter Array on the Atacama Cosmology Telescope}
 
\author{R. Datta \and J. Austermann \and J.A. Beall \and D. Becker \and K.P. Coughlin \and S.M. Duff \and P.A. Gallardo \and E. Grace \and M. Hasselfield \and S.W. Henderson \and G.C. Hilton \and S.P. Ho \and J. Hubmayr \and B.J. Koopman \and J.V. Lanen \and D. Li \and J. McMahon \and C.D. Munson \and F. Nati \and M.D. Niemack \and L. Page \and C.G. Pappas \and M. Salatino \and B.L. Schmitt \and A. Schillaci \and S.M. Simon \and S.T. Staggs \and J.R. Stevens \and E.M. Vavagiakis \and J.T. Ward \and E.J. Wollack}

\institute{R. Datta \and K.P. Coughlin \and J. McMahon  \and C.D. Munson \at Department of Physics, University of Michigan Ann Arbor, MI, USA 48109, \\\email{dattar@umich.edu} 
\and
J. Austermann \and J.A. Beall \and D. Becker \and S.M. Duff \and J. Hubmayr \and G.C. Hilton \and J.V. Lanen \and D. Li \at NIST Quantum Devices Group, 325 Broadway Mailcode 817.03, Boulder, CO, USA 80305
\and
P.A. Gallardo \and S.W. Henderson \and B.J. Koopman \and M.D. Niemack \and J.R. Stevens \and E.M. Vavagiakis \at Department of Physics, Cornell University, Ithaca, NY, USA 14853
\and
E. Grace \and S.P. Ho \and L. Page \and C.G. Pappas \and M. Salatino \and A. Schillaci \and S.M. Simon \and S.T. Staggs \at Department of Astrophysical Sciences, Peyton Hall, Princeton University, NJ USA 08544
\and
M. Hasselfield \at Joseph Henry Laboratories of Physics, Jadwin Hall, Princeton University,  NJ, USA 08544
\and
D. Li \at SLAC National Accelerator Laboratory, 2575 Sandy Hill Road, Menlo Park, CA 94025
\and
F. Nati \and B.L. Schmitt \and J.T. Ward \at Department of Physics and Astronomy, University of Pennsylvania, 209 South 33rd Street, Philadelphia, PA, USA 19104
\and
E.J. Wollack \at NASA Goddard Space Flight Center, Greenbelt, MD 20771 USA}


\date{the date of receipt and acceptance should be inserted later}

\maketitle

\begin{abstract}

We present the design and the preliminary on sky performance with respect to beams and pass-bands of a multichroic polarimeter array covering the 90 and 146 GHz Cosmic Microwave Background (CMB) bands and its enabling broadband optical system recently deployed on the Atacama Cosmology Telescope (ACT). The constituent pixels are feedhorn-coupled multichroic polarimeters fabricated at NIST. This array is coupled to the ACT telescope via a set of three silicon lenses incorporating novel broad-band metamaterial anti-reflection coatings. This receiver represents the first multichroic detector array deployed for a CMB experiment and paves the way for the extensive use of multichroic detectors and broadband optical systems in the next generation of CMB experiments.

\keywords{Antireflection Coating, Cosmic Microwave Background, Feedhorn, Millimeter-wave, Polarimeter, Silicon Lenses, Superconducting Detectors, TES}

\end{abstract}

\section{Introduction}
CMB polarization provides a unique window into the physics of inflation \cite{CMBPol:Bauman}, an alternative means to measure the neutrino mass sum \cite{CMBPol:Smith}, and enables a multitude of other astrophysical studies. Detections of the small angular scale B-mode polarization from  gravitational lensing \cite{Bmode:Lensing} have been reported \cite{SPT:Hanson, PB:bm, BICEP:Ade, ACTPol:vanEngelen} in the last two years. Recent results from Planck and BICEP \cite{Planck:BICEP} show that measurements of the inflationary B-mode signal at larger angular scales requires instruments capable of mapping the CMB at multiple frequency bands with unprecedented sensitivity and excellent control over systematics. Feedhorn-coupled detectors with sensitivity to multiple spectral bands within a single focal plane element offer an avenue to achieve these goals. 

The Atacama Cosmology Telescope Polarimeter (ACTPol) instrument \cite{ACTPol:Niemack} is optimized to make arcminute resolution measurements of CMB polarization, CMB lensing, and other secondary anisotropies. This paper describes the design and presents the preliminary on-sky performance with respect to passbands of the recently deployed array of multichroic polarimeters on the ACT telescope. In Sec. \ref{sec:desnperf} we describe the design of the multichroic array. In Sec. \ref{sec:actopt} we discuss the ACTPol optical design, including the broad-band anti-reflection (AR) coating of the large format silicon optics. On-telescope measurement of the detector passbands is presented in Sec. \ref{sec:ftsmeas}. We conclude in Sec. \ref{sec:conc} with a brief discussion of the ongoing work on the next generation upgrade to the ACTPol instrument, Advanced ACTPol (AdvACT). 

\section{Multichroic Pixel and Array Design}
\label{sec:desnperf}

\begin{figure}
\begin{center}
\includegraphics[width=1.0\linewidth,keepaspectratio]{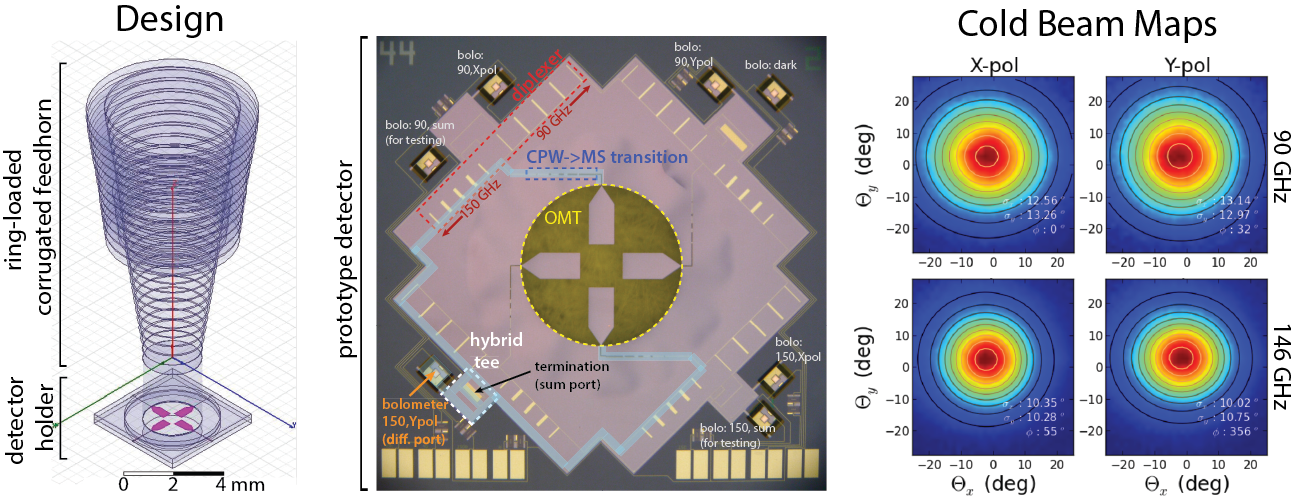}
\end{center}
\caption{{\it Left:} Design of a single horn coupled multichroic polarimeter. {\it Center:} Photograph of a prototype 90/146 GHz multichroic detector with the major components labeled. For clarity, the path light follows to reach the bolometer corresponding to Y polarization in the 146 GHz band has been highlighted. For testing purposes, the prototype pixel included additional bolometers connected to the hybrid tee sum port. {\it Right:} The prototype detector beams as measured by raster scanning with a hot thermal source in the lab at NIST. (Color figure online)}
\label{fig:pixdesign}
\end{figure}

Figure \ref{fig:pixdesign} shows a design of a single multichroic polarimeter with sensitivity to the 90 and 146 GHz CMB bands, photograph of a prototype detector chip, and the measured beams. The beam pattern is defined by a broad-band ring-loaded corrugated feedhorn \cite{Ringloadedguide} which couples light onto a planar detector chip that sits above a waveguide back short to prevent leakage of electromagnetic fields. A broad-band Ortho-Mode Transducer (OMT) \cite{multichroic:McMahon} separates the incoming radiation according to linear polarization and couples it to high impedance Co-Planar Waveguide (CPW) lines. The signal transitions onto low impedance Micro-Strip (MS) lines via a broad-band CPW to MS transition. Next, diplexers comprised of two separate five pole resonant stub band-pass filters split each signal into two frequency channels. The signals from opposite OMT probes within a single frequency band are then combined onto a single MS line using the difference output of a hybrid tee \cite{coupler:Knoechel}. These signals are detected using Transition Edge Sensor (TES) bolometers and read out using SQUIDs. Signals appearing at the sum output of the hybrid are routed to a termination resistor and discarded. This architecture described above offers excellent control over beam systematics provided by corrugated feeds, a frequency independent polarization axis, and a metal skyward aperture to minimize electrostatic buildup that is useful for future space applications \cite{CMB:Space}. 

\begin{figure}
  \begin{center}
    \includegraphics[width=1.0\textwidth]{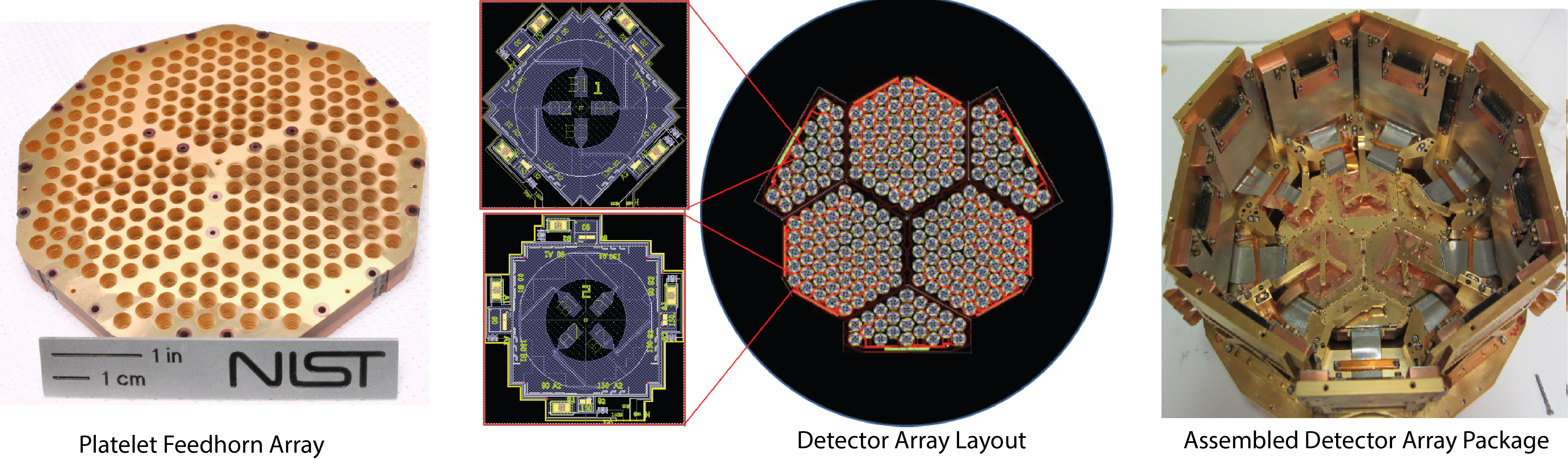}
    \caption{{\it Left:} Photograph of the gold plated silicon platelet feedhorn array. {\it Center:} Multichroic detector array layout composed of three $3^{\prime\prime}$ diameter hex sub-arrays (61 pixels each) and three $3^{\prime\prime}$ diameter semi-hex sub-arrays (24 pixels each). The full array consists of 255 pixels equally split into two orientations differing by 45 degrees. Each pixel orientation is uniformly distributed across the full array. {\it Right:} Fully assembled detector array package viewed from the non-sky side. (Color figure online)}
\label{fig:array}
  \end{center}
\end{figure}

Measurements of the beam, pass-band, and optical efficiency presented in \cite{multichroic:Datta} demonstrated the performance of early prototype detectors. This pixel design was incorporated into a detector array (see Figure \ref{fig:array}) that was designed to use the existing ACTPol 1024 channel detector readout. This choice leads to near optimal sensitivity at 90 GHz, but is sub-optimal for 146 GHz. With AdvACT, we will field new arrays with more detectors optimized to maximize the combined sensitivity. The full multichroic array consists of 255 pixels with a total of 1020 polarization sensitive bolometric detectors fed by a 140 mm diameter silicon platelet monolithic feedhorn array \cite{Truce:britton}. It was deployed in January 2015, as the third and final ACTPol array, along with the two existing 146 GHz single frequency arrays.

\section{Multichroic ACTPol Optics}
\label{sec:actopt}

\begin{figure}
  \begin{center}
    \includegraphics[width=1.0\textwidth]{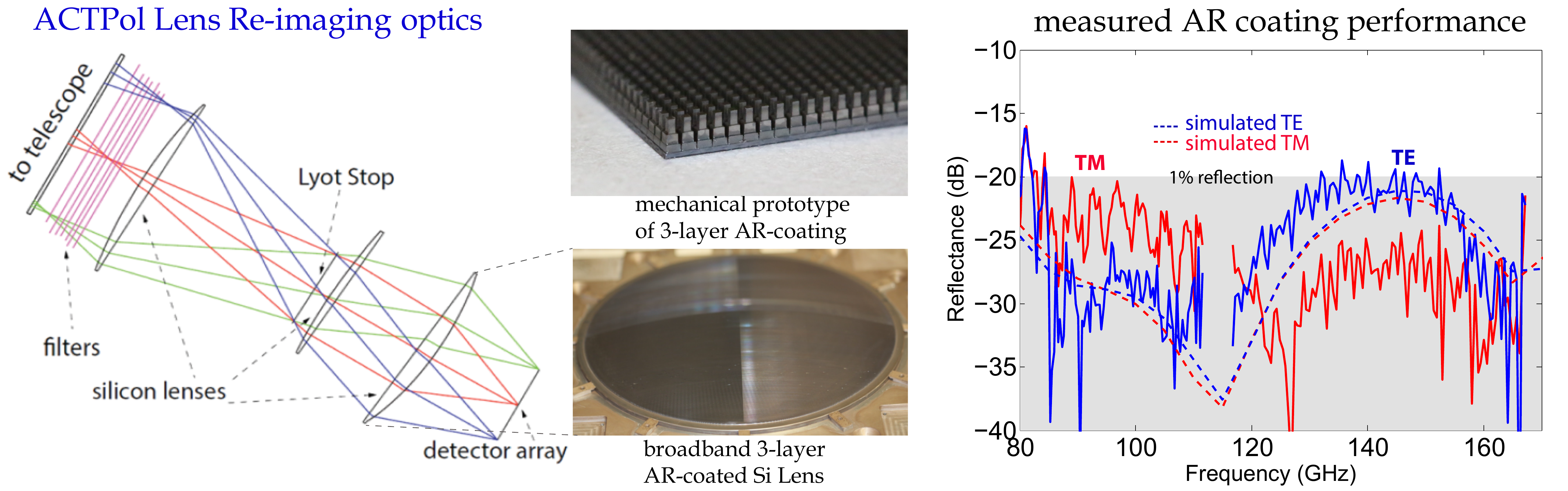}
  \end{center}
  \caption{{\it Left:} A ray diagram of the multichroic ACTPol re-imaging optics, which includes three silicon lenses feeding a detector array. {\it Center:} Photographs of a mechanical prototype of the 3-layer AR coating and one of the AR-coated lens. The geometry of the coating consists of square pillars that are 185, 325, 420 $\mu$m wide and 470, 315, 245 $\mu$m deep with a spacing of 450 $\mu$m. These three layers correspond to effective indices of 1.28, 1.95, and 2.84 respectively. {\it Right:} Reflection from the surface of an AR-coated lens measured with a reflectometer set up for both linear polarizations (TE and TM modes). The dashed lines are from simulations at $15^\circ$ angle of incidence. (Color figure online)} 
\label{fig:ARC}
\end{figure}

The multichroic ACTPol array is coupled to the telescope via three cryogenic silicon lenses (see Figure \ref{fig:ARC} $\it{left}$) similar to those used for ACTPol \cite{ACT:Swetz, ACT:Fowler}. The vacuum window at the Gregorian focus of the telescope (focal ratio of 2.5) is made of $\sim$6.9 mm thick Ultra-High Molecular Weight Polyethylene (UHMWPE) with an optimized AR coating consisting of bonded expanded-Teflon sheets $\sim$0.43 mm thick glued to the front and back surfaces of the UHMWPE. A series of low-pass (LP) capacitive-mesh filters placed at successive cryogenic stages along the optical path limit loading on the next stage. A series of infrared-blocking (IR) reflective filters \cite{filter:Tucker} mitigate heating of the center of the poorly conducting LP filters and consequently reduce the radiative loading from the windows. Cryogenic baffles at different temperature stages limit spillover inside the dewar. The cold silicon optics (see Figure \ref{fig:ARC} $\it{middle}$), consisting of three cryogenic silicon lenses, re-image the Gregorian focus onto the focal plane detector array. The first lens, located just after the Gregorian focus, forms an image of the primary mirror on a cold (1K) lyot stop that defines the illumination of the secondary and primary reflectors. The second and third lenses refocus the sky onto the focal plane arrays. Due to the non-telecentric, off-axis Gregorian design of the telescope, the optimal focal plane is not perpendicular to the optical axis.    

Implementing multichroic arrays requires optical designs with wide bandwidth. The limiting element for the ACTPol optics in terms of loss is the AR coating for the three silicon lenses that make up the cryogenic re-imaging optics. We have developed an approach to AR coating cryogenic silicon lenses with metamaterials fabricated by cutting sub-wavelength features into the surfaces using silicon dicing saw blades \cite{lenses:Datta}. The AR coating optimized for 75 - 170 GHz frequency band consists of three layers of square pillars, which ensures polarization symmetry. This coating reduces reflections to less than 0.5\% averaged over each of the required bands. The right panel of figure \ref{fig:ARC} shows the measured reflection from one of these AR coated lenses and a photograph of a mechanical prototype. The measured performance differs from our predictions at the sub-percent level. This is due to imperfections of the machining process we identified after fabricating this first batch of lenses.  We have developed a new approach to improve this for AdvACT.  

\section{On-telescope Performance}
\label{sec:ftsmeas}

\begin{figure}
\begin{center}
\includegraphics[width=1.0\linewidth,keepaspectratio]{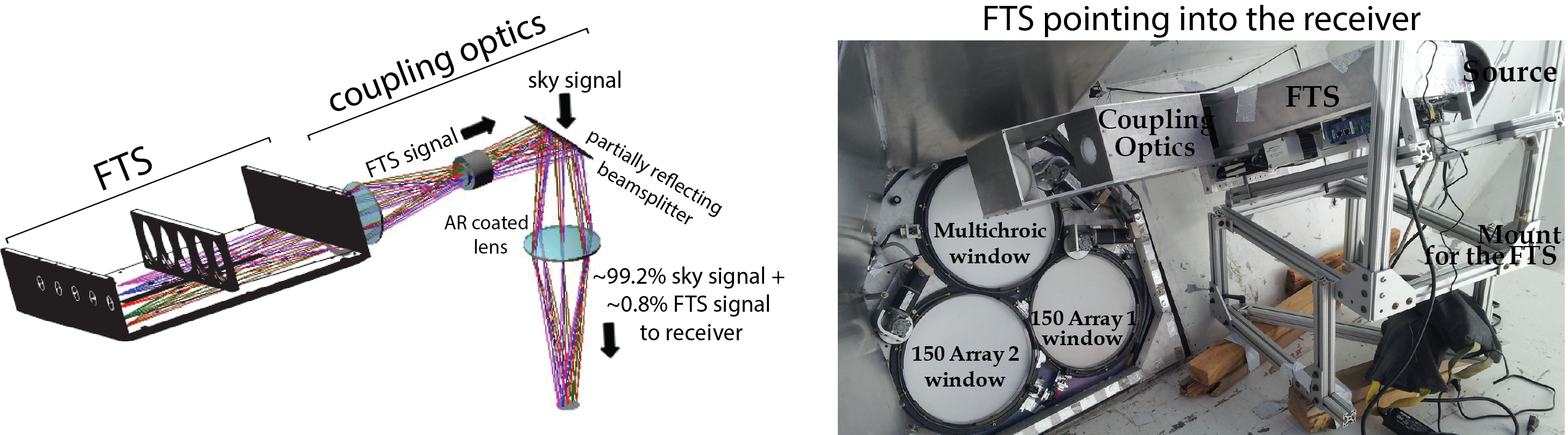}
\end{center}
\caption{{\it Left:} A schematic of the optics which couple the FTS output port to the ACTPol receiver. {\it Right:} A photograph of the FTS setup in front of the ACTPol receiver to measure the instrument passbands. The two different step size and throw combinations of the dihedral mirrors are (0.03~mm, 100~mm) for higher resolution scanning and (0.15~mm, 20~mm) for lower resolution, high bandwidth scanning. (Color figure online)}
\label{fig:ftssetup}
\end{figure}

The passband of the multichroic receiver was measured after its deployment on the ACT telescope. Interferograms were acquired using a symmetric Fourier Transform Spectrometer (FTS) using a Martin-Puplett type interferometer \cite{pixie:FTS} operating in a step-and-integrate mode. In this mode, the optical path difference in the interferometer is incremented in discrete steps, and the detector signal is integrated when the interferometer mirrors are stationary. A programmable linear stage shifts the position of back-to-back dihedral mirrors in the optical path to introduce a phase delay. One input port was terminated by a beam-filling source, either a 600~C thermal blackbody or a sheet of eccosorb immersed in liquid nitrogen at 77~K. The additional ports were covered with 300~K eccosorb to help damp spurious reflections. The FTS was calibrated using a microwave transmitter in a frequency range between 80 and 170 GHz. The step-and-integrate mode requires an additional modulation which was implemented by chopping the source at 7~Hz so that the detector alternately views source and ambient temperature. A coupling optics arm (see Figure \ref{fig:ftssetup}) comprised of three thick polyethylene (HDPE) lenses and a thin polyethylene (LDPE) beam splitter coupled the output of the FTS to the detectors, illuminating $\sim$5 pixels at a time. This coupling system (see Figure \ref{fig:ftssetup} $\it{left}$) was designed to minimize the emissive loading on the receiver from the plastic lenses. Its focal ratio of 1.6 ensures that the output can fill the beam of any of the feedhorns on the non-telecentric focal plane without having to point the FTS at an angle. The LDPE beam splitter reflects $<$1\% of the incident light which was focused on to the detectors. The remaining $>$99\% of the light passes through the beam splitter and reflects off the telescope onto the sky. This approach prevented the detectors from saturating and maintained $\sim$20 detectors, within the field of view of the FTS, in transition. The lens closest to the receiver window was AR coated by cutting concentric circular grooves on both surfaces. We measured the bandpasses of 5\% of the  $\sim$1000 detectors with $\sim$5 re-pointings (see Figure \ref{fig:ftssetup} $\it{right}$) of the FTS. We used two types of scans, one at high resolution across the nominal bands to measure their shapes, and another with low resolution across a wider band to check for out-of-band leakage. Data were stored with a sampling rate of 400 Hz over a 1 second integration time. The modulated signal was detected with the standard software lock-in technique. The interferograms were phase corrected and Fourier transformed to get the spectral response. The final spectra were corrected for the transmission of the lenses constituting the coupling optics, the reflection from the LDPE beam splitter, and the source blackbody spectrum.   

\begin{figure}
  \begin{center}
    \includegraphics[width=1.0\textwidth]{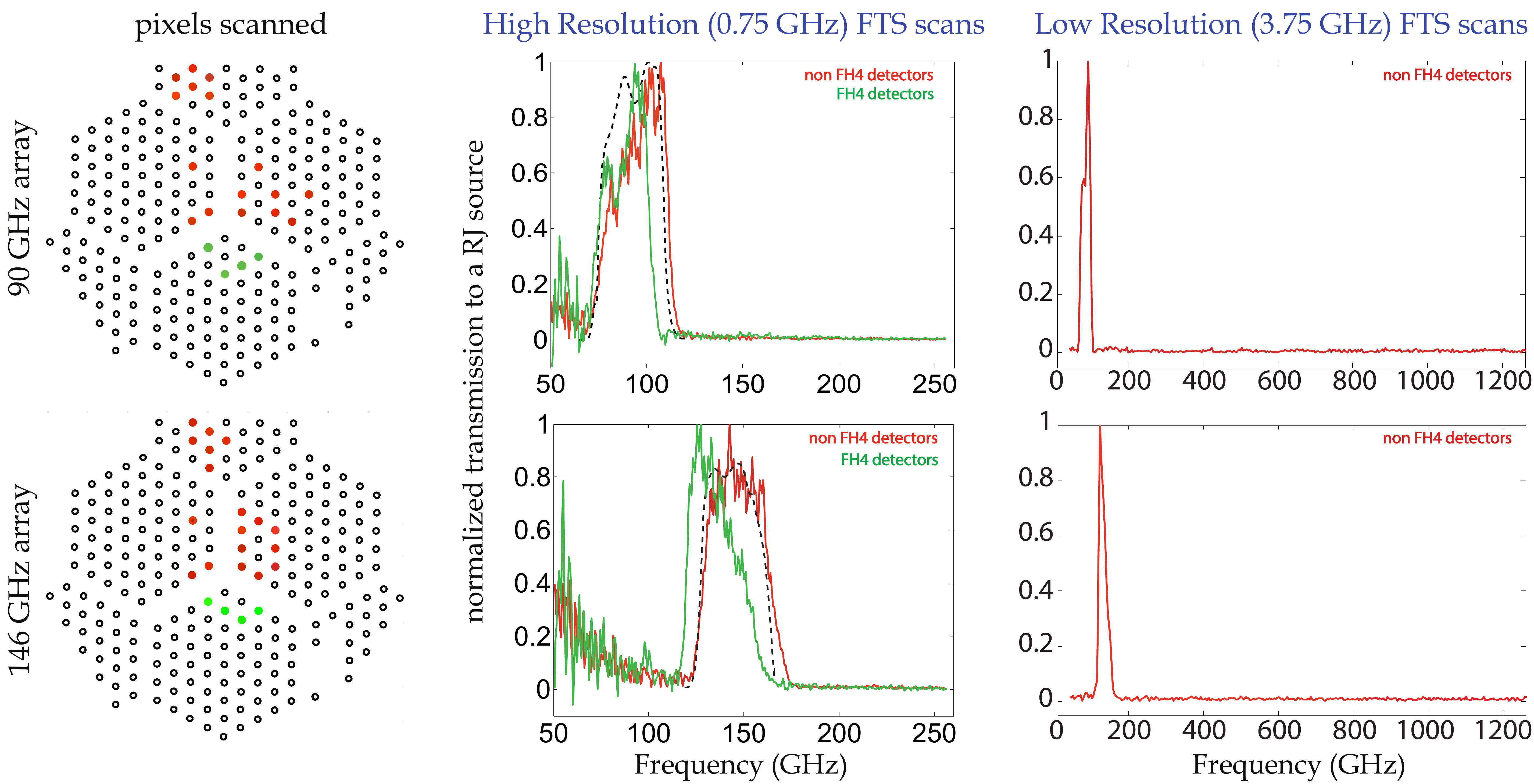}
    \caption{The left column shows schematics of the multichroic array, with each circle representing a pixel and the filled circles representing the detectors whose measured passbands are averaged and plotted in each row. The subsequent columns show averaged high resolution and low resolution spectra. The detectors on the bottom hex wafer, FH4, have passbands shifted down in frequency. The lowest band edge, defined by the feedhorn waveguide section, is not affected by this. The dotted lines are from simulations of the multichroic pixel, including the feedhorn and the OMT. No evidence for out-of-band leakage was found. (Color figure online)} 
\label{fig:spectra}
  \end{center}
\end{figure}

Figure \ref{fig:spectra} shows the measured spectra averaged over individual detectors in the instrument. We found that the detectors from one of the hex wafers had shifted passbands that could be explained by changes in dielectric properties. Using the measured spectral response, we follow the method described in \cite{WMAP:Lyman} to calculate the effective central frequency of broadband sources, including the CMB and the Sunyaev-Zeldovich (SZ) effect \cite{Dunkley:2012} and the effective bandwidth of the detectors. These are listed in Table 1. More details about the on-sky performance of the multichroic array are presented in \cite{LTD:Ho} in these proceedings, including detector sensitivities and time constants.

\begin{table}[t]
\begin{center}
\captionsetup{justification=centering}
\caption{Multichroic array effective central frequencies and bandwidth}
\begin{tabular}{ccc}
\hline
\hline
Array & 90 (GHz) & 146 (GHz) \\
\hline
Effective bandwidth$^{a}$ & 38$\pm$2 & 41$\pm$2\\
\hline
\hline
Effective band centers \\
\hline
Synchrotron$^{b}$ & 91.7$\pm$1.7 & 143.6$\pm$2.4\\
Free-free$^{b}$ & 92.5$\pm$1.7 & 144.2$\pm$2.4\\
Rayleigh-Jeans$^{b}$ & 95.3$\pm$1.7 & 146.1$\pm$2.4\\
Dusty source$^{b}$ & 97.3$\pm$1.7 & 147.7$\pm$2.4\\
CMB$^{b}$ & 97.1$\pm$1.7 & 146.6$\pm$2.4\\
SZ effect$^{b}$ & 94.1$\pm$1.7 & 143.7$\pm$2.4\\
\hline
\hline
Conversion factors \\
\hline
$\delta T_{CMB}/ \delta T_{RJ}$$^{c}$ & 1.27$\pm$0.04 & 1.69$\pm$0.05\\
$\delta W/ \delta T_{RJ} (pW/K)$$^{d}$ & 0.52$\pm$0.03 & 0.57$\pm$0.03\\
\hline\end{tabular}
\end{center}
$^{a,}$$^{b}$ These values are obtained by averaging over individual detectors (excluding the ones with the shifted passbands). The uncertainties are obtained from a combination of the variance of the measurements and an estimate of systematic error. \\
$^{b}$ Spectral index assumed: Synchrotron emission ($\alpha = -0.7$), Free- free emission ($\alpha  = -0.1$), Rayleigh-Jeans emission ($\alpha = 2.0$), Dusty source emission ($\alpha = 3.7$). \\
$^{c}$ This gives the conversion factor to RJ. \\
$^{d}$ This gives the power in a single mode of radiation. 
\label{tab:eff}
\end{table}

\begin{figure}
\begin{center}
    \includegraphics[width=0.8\textwidth]{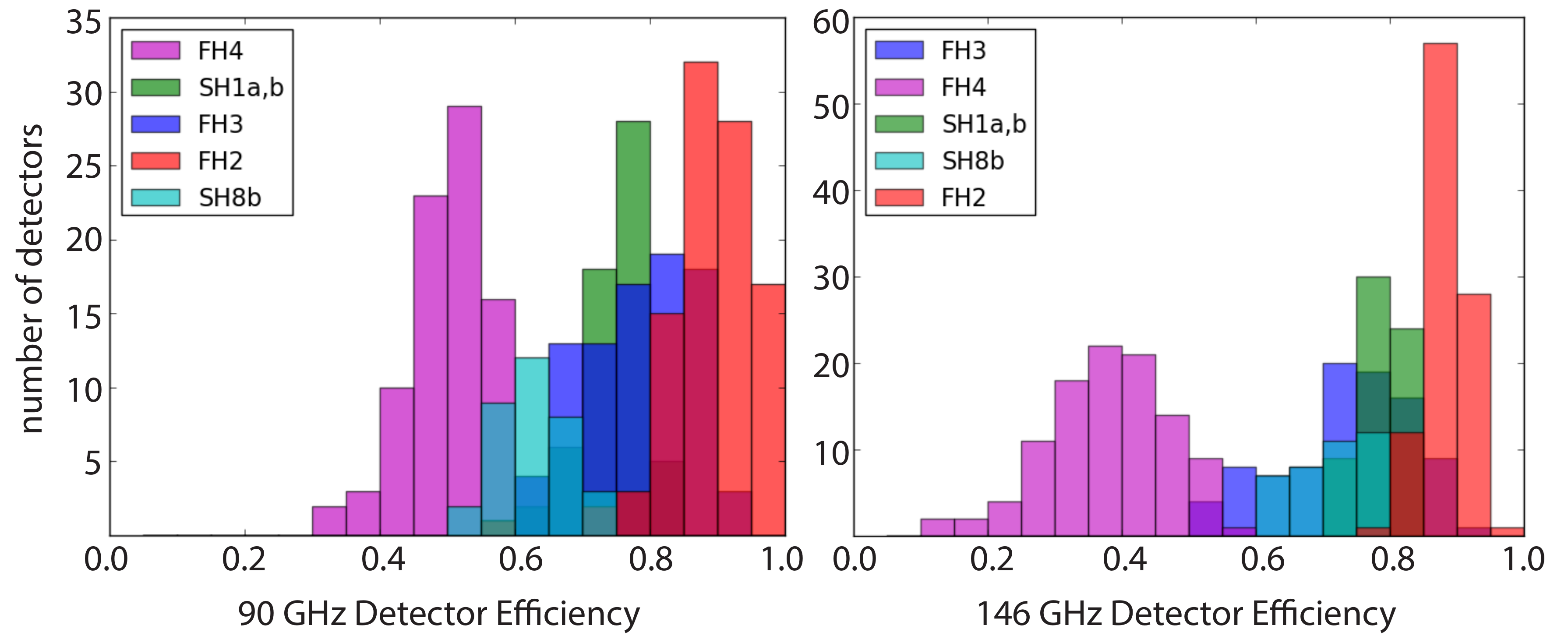}
    \end{center}
    \caption{A histogram plot of the 90 and 146 GHz detector efficiencies, relative to the simulated passband shown in dotted lines in Fig. 5. FH4, FH3 and FH2 represent the three hex sub-arrays and SH1a, SH1b and SH8b represent the three semi-hex sub-arrays. (Color figure online)} 
\label{fig:opteff}
\end{figure}

Figure \ref{fig:opteff} shows the distribution of the efficiencies of the 90 and 146 GHz detectors in the deployed multichroic array, with respect to the expected passbands from simulation of the pixel and the feedhorn. These are computed using the relative response of the detectors to the atmosphere. The low optical efficiency of the FH4 detectors can be explained by their shifted bands relative to the simulated design.

\section{Conclusion and Future Work}
\label{sec:conc}

We have developed and deployed a polarization sensitive multichroic array with 255 feedhorns and 1020 bolometers on ACT in Chile. The preliminary sensitivity of the array is $<$10~\rm{$\mu$K}$\sqrt{s}$, with the exact value changing with the atmospheric loading, which makes it the most sensitive ACTPol array. \hbox{AdvACT} \cite{LTD:Henderson} will be a staged upgrade over three years of ACTPol's three existing detector arrays and their optics. The new detector arrays \cite{LTD:Duff,LTD:Li} under development will be more densely packed multichroic arrays spanning from 28 GHz to 230 GHz with an upgraded read-out system.

\begin{acknowledgements}
This work was supported by NASA through awards NNX13AE56G and NNX14AB58G and by the U.S. National Science Foundation through awards AST-0965625 and PHY-1214379. The NIST authors would like to acknowledge the support of the NIST Quantum Initiative. The work of KPC, EG, BJK, BLS, CDM, JTW and SMS were supported by NASA Office of the Chief Technologists Space Technology Research Fellowship awards.
\end{acknowledgements}

\end{document}